\documentclass[twocolumn,showpacs,preprintnumbers,amsmath,amssymb,superscriptaddress]{revtex4}
\usepackage{amssymb}
\usepackage{graphicx,color}
\usepackage{bm}

\usepackage{graphicx} \usepackage{dcolumn} \usepackage{bm}
\newcommand{\beq}{\begin{equation}}\newcommand{\eeq}[1]{\label{#1}
\end{equation}}\newcommand{\beqar}{\begin{eqnarray}}\newcommand{\eeqar}[1]
{\label{#1}
\end{eqnarray}}\newcommand{\bmath}{\begin{displaymath}}\newcommand{\emath}{\end{displaymath}}\newcommand{\bitem}{\begin{itemize}}\newcommand{\eitem}{\end{itemize}}
 
\begin{document}

\title{\Large \bf  Strong longitudinal color-field effects
     in \\ $pp$ collisions at energies available at the CERN 
     Large Hadron Collider.}

\newcommand{\mcgill}{McGill University, Montreal, Canada, H3A 2T8}

\newcommand{\columbia}{Columbia University, New York, N.Y. 10027}

\affiliation{\mcgill}
\affiliation{\columbia}

\author{~V.~Topor~Pop} \affiliation{\mcgill}
\author{~M.~Gyulassy} \affiliation{\columbia} 
\author{~J.~Barrette} \affiliation{\mcgill}
\author{~C.~Gale} \affiliation{\mcgill}
\author{~A.~Warburton} \affiliation{\mcgill}

\date{January 25, 2011}

\begin{abstract}

We study the effect of  strong longitudinal color fields (SCF)
in  $p+p$ reactions up to Large Hadron Collider energies in the framework of
the {\small HIJING/B\=B } v2.0 model that combines (collinear factorized) pQCD
multiple minijet production with soft longitudinal string excitation and
hadronization. The default vacuum string tension, $\kappa_0$ = 1 GeV/fm, 
is replaced by an effective energy dependent string tension, 
$\kappa(s) = \kappa_0 (s/s_0)^{0.06}$ that increases monotonically with 
center-of-mass energy. The exponent $\lambda=0.06$ is found
sufficient to reproduce well the energy
dependence of multiparticle observables in RHIC, Tevatron, as
well as recent LHC data. This exponent is half of that predicted by the 
Color Glass Saturation (CGC) model, $\lambda_{CGC}=0.115$, where
gluon fusion multiparticle production mechanisms are assumed.
In {\small HIJING/B\=B} v2.0, the rapid growth
of $dN_{ch}/d\eta$ with energy is due to the interplay of copious minijet
production with increasing of strong color field (SCF) contributions.
The large (strange)baryon-to-meson ratios measured at  Tevatron energies
are well described. A significant enhancement of these
ratios is predicted up to the highest LHC energy (14 TeV). The effect of J\=J
loops and SCF on baryon-anti-baryon asymmetry, and its relation 
to baryon number transport, is also discussed.

\end{abstract}


\maketitle




\section{Introduction}

With the commissioning of the Large Hadron Collider (LHC), it 
will soon be  
possible to test models of multiparticle production in hadron-hadron 
collisions up to an energy of 14 TeV.
Charged particle densities, $dN_{\rm ch}/d\eta$, especially
the values at mid-rapidity   
and their dependence on center-of-mass energy $\sqrt{s}$ (c.m.s.) 
are important for understanding
the mechanism of hadron production and the interplay of soft and hard
scattering contributions in the LHC energy range.
The rate of parton-parton and multi parton-parton (MPI) 
scattering are strongly correlated to the
observed particle multiplicity (related also to
{\em initial entropy} and {\em initial energy density} generated in the
collision process).
New data on inclusive charged particle distributions from the LHC in $pp$
collisions have become available 
\cite{Khachatryan:2010xs,Khachatryan:2010us,alice:2009dt,Aamodt:2010my,Aamodt:2010pp,Aamodt:2010ft,Aad:2010rd,atlas_conf24,atlas_conf31,atlas_conf46,atlas_conf47,Collaboration:2010ir}.  
These results complement previous data on $pp$ and $p\bar{p}$
collisions taken at lower energies $\sqrt{s}$ = 0.02 - 1.96 TeV 
\cite{:2008ez,cdf2009_prd79,e735_plb94,ua1_npb90,cdf1990_prd41,ua5_zfurp_c43,ua5_zfurp_c37,cdf1988_prl61,ua5_prep_154,ua5_zfurp_c33,npb_129_1977}.
Many of these measurements 
have been used to constrain phenomenological models of soft-hadronic 
interactions and to predict properties at higher energies
\cite{armesto_jpg08,armesto_ar09,Mitrovski:2008hb,McLerran:2010ex,Levin:2010dw,Levin:2010zy,merino1_2010,Skands:2010ak,buckley1_2009,moraes1_2007,reygers_2010,Kraus:2008fh,Kaidalov:2009hn,Sarkisyan:2010kb}.
The recent LHC data may lead to a better theoretical
understanding based on a Quantum Chromodynamics (QCD) approach 
\cite{Warburton:2010bj,Seymour:2010ih,Sassot:2010bh,Deng:2010mv}.

The Heavy Ion Jet Interacting Generator ({\small HIJING})
\cite{mik_wang_94} and {\small HIJING/B\=B} v1.10 models 
\cite{miklos_99} have been used extensively to study 
particle production in $pp$ collisions and 
to determine the physical properties of the ultra-dense matter produced in 
relativistic heavy-ion collisions.
In the LUND \cite{Andersson:1986gw}
 and Dual-Parton (DPM) \cite{Capella:1992yb} models multi QCD 
strings or flux tubes  
were proposed to describe soft multiparticle production
in longitudinal color fields. Color exchange between high $x$ partons in
the projectile and target create confined color flux tubes of tension
($\approx 1$ GeV/fm), that must
neutralize through pair production or color singlet hadronization
approximately uniformly in rapidity.
In nucleus-nucleus collisions, the $A^{1/3}$ enhancement of the local
parton density of high $x$ partons allows for higher color Casimir
representations to be excited.  Those
flux tubes with stronger longitudinal color fields 
than in average $pp$ reactions have been
called color ropes \cite{Biro:1984cf} and naturally have higher
string tensions \cite{Gyulassy:1986jq}. Recently, an extension
of Color Glass Condensate (CGC) theory has proposed a more detailed
dynamical ``GLASMA'' model \cite{larry_2009,mclerran_08} of color ropes. 
 
In the {\small HIJING} model \cite{mik_wang_94} 
the soft beam jet fragmentation is modeled by simple diquark-quark strings 
as in the LUND model with multiple 
 gluon kinks induced by soft gluon radiations. 
Hard collisions are included with standard perturbative QCD
(pQCD) as programmed in the  PYTHIA generator \cite{Sjostrand:1993yb}.
However, HIJING differs from PYTHIA by 
the incusion of a geometric scaling multiple jet production model. 
Thus this model contains both longitudinal field induced
 {\em soft} beam jet multiparticle production and collinear factorized
pQCD based {\em hard} multiple
jet production for  $p_T\ge p_0=2$ GeV/c.

A systematic comparison with data on $pp$ and $p\bar{p}$ 
collisions in a wide energy range \cite{mik_wang_94} 
revealed that minijet production and fragmentation as
implemented in the {\small HIJING} model provide a simultaneous and 
consistent explanation of several effects: the inclusive spectra at moderate 
transverse momentum ($p_T$); the energy dependence of the central
rapidity density; the two particle correlation function;  
and the degree of violation of 
Koba-Nielsen-Olesen (KNO) scaling \cite{wang1_91}, \cite{wang2_92} up
to Tevatron energy ($\sqrt{s} = 1.8$ TeV).
However, the model failed to describe the dependence of the mean value 
of transverse momentum ($<p_T>$) on charged particle 
multiplicity ($N_{\rm ch}$). Wang argued \cite{wang2_92} 
that by requiring high $N_{\rm ch}$ within a limited 
pseudorapidity ($\eta$) range
one necessarily biases the data towards higher $p_T$ minijets,
hence the observed increase of $<p_T>$ versus (vs.) $dN_{\rm ch}/d\eta$
\cite{wang2_92}.
This effect has also been associated with the presence of transverse
flow of the hadronic matter \cite{plevai_91,Pierog:2010wa}
and was proposed as possibly due to 
 {\em quark-gluon plasma} (QGP) formation already in $pp$ collisions.

Initial states of color gauge fields produced in high-energy 
heavy-ion collisions have also recently been discussed 
in Ref.~\cite{Iwazaki:2009cf}. Decay of a strong color electric field (SCF) 
($E > E_{\text critical}$=$10^{18}$ V/m) due to the 
Schwinger mechanisms \cite{schwinger} plays an 
important role at the initial stage of heavy-ion collisions at 
ultra-relativistic energies. 
A thermalization scenario based on the analogy between Schwinger mechanisms
and the Hawking-Unruh effect has been proposed \cite{kharzeev_07}.
It was also suggested that the back-reaction and screening effects 
of quark and anti-quark pairs on external electric field could even lead to the
phenomenon of plasma oscillations 
\cite{Tanji:2008ku,Ruffini:2009hg,Tanji:2010eu}.  

Recently, the Schwinger mechanism has been revisited \cite{cohen_jul08} 
and pair production in time-dependent 
electric fields has been studied \cite{gies_jul08}.
It was concluded that particles with large momentum were likely to have 
been created earlier, and for very short temporal widths 
($\Delta \tau \approx 10 {\rm t}_{\rm c}$, where the Compton time 
${\rm t}_{\rm c}=1/{\rm m}_{\rm c}$) and as a consequence 
the  Schwinger formula could 
underestimate the reachable particle number density.
In previous papers,  
we have shown that the dynamics of strangeness production in $pp$ and 
Au + Au collisions at Relativistic Heavy Ion Collider (RHIC) energies 
deviates considerably from calculations based on Schwinger-like
estimates for homogeneous and constant color fields  
and point to the contribution of fluctuations of transient strong 
color fields (SCF) \cite{prc72_top05,prc75_top07,armesto2_08,top_prl2009}.

In this paper we extend our study of the dynamic consequences of
SCF in the framework of the  
{\small HIJING/B\=B v2.0} model \cite{prc75_top07}
to particle production in hadron-hadron collisions at LHC energies.  
We explore dynamical effects associated with
long range coherent fields ({\it i.e}, strong longitudinal color fields, SCF),
including baryon junctions and loops \cite{prc72_top05,top04_prc}, 
with emphasis on the new observables measured in $pp$ collisions at
the LHC by the ALICE, ATLAS and CMS collaborations.
Our study aims to investigate a broad set of observables 
sensitive to the dynamics of the collisions, 
covering both longitudinal and transverse degree of freedom.
In addition, this study is intended to provide the $pp$ baseline to
future extrapolations to LHC studies of
proton-nucleus ($p + A$) and nucleus-nucleus ($A + A$) collisions.

\section{Outline of {\small HIJING/B\=B v2.0} model.}

In  {\small HIJING/B\=B v2.0},  
in addition to conventional quark-diquark longitudinal electric fields,
novel color flux topology junction anti-junction (J\=J) loops 
are also implemented.
In a dual superconductor model of color confinement
for the three-quark positioning in a Y - geometry, 
the flux tubes converge first toward the
center of the triangle and there is also another component
that runs in the opposite direction (like a {\em Mercedes star})
\cite{Ripka:2003vv}.
Unlike the conventional diquark-quark  implemented  
in LUND and the HIJING model \cite{mik_wang_94}, the
 {\small HIJING/B\=B} v1.10 \cite{miklos_99} model allows 
 the diquark-quark to split with the
three independent flux lines tied together with an $\epsilon_{ijk}$
a junction and terminating on three delocalized fundamental Casimir quarks.
We introduced \cite{top04_prc} 
a new version (v2.0) of {\small HIJING/B\=B} that differs from 
{\small HIJING/B\=B} v1.10 \cite{miklos_99}  
in its implementation of additional more complex flux topologies
via junction anti-junction (J\=J) loops. 
We parametrize the probability
that a junction loop occurs in the string. Moreover, we enhance 
the intrinsic (anti)diquark-quark $p_T$ kick (by a factor $f$ = 3)
of all ($q$-qq) strings that 
 contain one or multiple J\=J loops. The reason for this is the
mechanism behind the dynamic of diquark-quark breaking 
(see Ref.~\cite{prc75_top07} for details).

In string fragmentation phenomenology, it has been proposed
that the observed strong enhancement of strange particle
production in nuclear collisions
could be naturally explained via strong color field (SCF) 
effects \cite{Gyulassy:1986jq} .
For a uniform chromoelectric flux tube with field ({\it E}), 
the pair production rate \cite{Biro:1984cf, Gyulassy:1986jq,cohen_jul08}  
per unit volume for a (light)heavy quark ($Q$) is given by
\begin{equation}
\Gamma =\frac{\kappa^2}{4 \pi^3} 
{\text {exp}}\left(-\frac{\pi\,m_{Q}^2}{\kappa}\right),
\label{eq:prod_rate}
\end{equation}
where $Q={\rm qq}$ (diquark), $s$ (strange), $c$ (charm) or $b$
(bottom). The {\em current 
quark masses} are $m_{\rm qq}$ = 0.45 GeV \cite{ripka:2005},  
$m_s=0.12$ GeV, $m_{\rm c}=1.27$ GeV, and $m_b=4.16$ GeV \cite{pdg:2010}.
The {\em constituent quark masses} of light non-strange quarks 
are $M_{u,d}$ = 0.23 GeV, of the strange quark is $M_s$=0.35 GeV 
\cite{armesto2001}, and of the diquark is $M_{\rm qq}=0.55 \pm 0.05$ GeV 
\cite{ripka:2005}.

Note that $\kappa=|eE|_{eff}= \sqrt{C_2(A)/C_2(F)}\, \kappa_0$ 
is the effective string tension in low energy $pp$ reactions
with  $\kappa_0 \approx 1 $ GeV/fm
and $C_2(A)$, $C_2(F)$ are the second order Casimir operators 
(see Ref. \cite{Gyulassy:1986jq}).
An enhanced rate for spontaneous pair production is naturally
associated with 
``{\em strong chromo-electric fields}'', such 
that $\kappa/m_{\rm Q}^2\,\,>$ 1 {\em at least some of the time}.
In a strong longitudinal color electric field, 
the heavier flavor suppression factor 
$\gamma_{Q\bar{Q}}$ varies with string tension 
via the well known Schwinger formula \cite{schwinger}, 
since
\begin{equation}
\gamma_{Q\bar{Q}} = \frac{\Gamma_{Q\bar{Q}}}{\Gamma_{q\bar{q}}} =
{\text {exp}} \left(-\frac{\pi(M_{Q}^2-m_q^2)}{\kappa_0} \right)
 < 1
\label{eq:gamma_supress}
\end{equation}
for $Q = {\rm qq}$, $s$, $c$ or $b$ and $q = u$, $d$.
In our calculations, we assume that $M^{\rm eff}_{qq}$ = 0.5 GeV, 
$M^{\rm eff}_{s}$ = 0.28 GeV, $M^{\rm eff}_{c}$ = 1.30 GeV.
Therefore, the above formula implies a 
suppression of heavier quark production according to
$u$ : $d$ : ${\rm qq}$ : $s$ : $c$ $\approx$ 1 : 1 : 0.02 : 0.3 : 10$^{-11}$ 
for the vacuum string tension $\kappa_0$ = 1 GeV/fm.  
For a color rope, on the other hand,
if the {\em average string tension} value $\kappa$ 
increases, the suppression factors $\gamma_{Q\bar{Q}}$ increase.
We show below that this dynamical mechanism improves considerably the 
description of the strange meson/hyperon 
data at the Tevatron and at LHC energies.

Saturation physics is based on the observation
that small-x hadronic and nuclear wave functions, and, thus the 
scattering cross sections as well, are described by the same internal 
momentum scale known as the {\em saturation scale} ($Q_{\rm sat}$).
A recent analysis of $pp$ data up to LHC 7 TeV
has shown that, with  the $k_T$ factorized (GLR) gluon fusion approximation
\cite{Gribov:1984tu},
the growth of the $dN_{ch}/d\eta$ can be accounted for
if the saturation scale grows with c.m.s. energy as
\begin{equation}
Q_{\rm sat}^2(s) = Q_0^2 (s/s_0)^{\lambda_{\rm CGC}},
\label{eq:larry10}
\end{equation}
 with $\lambda_{\rm CGC} \approx 0.115$. 
The saturation scale is also increasing with atomic number as $A^{1/6}$
\cite{kharzeev_07}. It was argued that the effective 
string tension ($\kappa$) of color ropes should scale with
$Q_{\rm sat}^{2}$ \cite{kharzeev_07,Tanji:2008ku}. 

\begin{figure} [h!]
\centering
\includegraphics[width=0.9\linewidth,height=6.0cm]{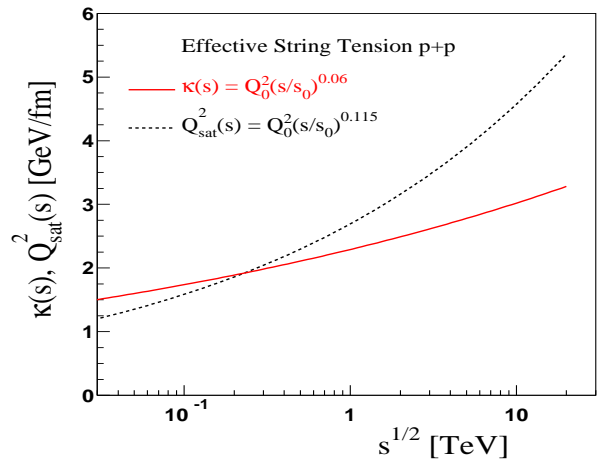}
\vskip 0.5cm\caption[energy dependence of string tension] {\small 
(Color online) 
Energy dependence of the effective string tension $\kappa(s)$ in $pp$
collisions (Eq.~\ref{eq:kappa_sup}). 
The values $Q_{\rm sat}^2$ (Eq.~\ref{eq:larry10}) are from CGC model
fits to LHC data from Ref.~\cite{McLerran:2010ex}. 
\label{fig:kappa_vs_cgc}
}
\end{figure}

However, in HIJING
the string/rope fragmentation is the only soft source of multiparticle
production and multiple minijets provide a semi-hard additional
source that is computable within collinear factorized standard pQCD
with initial and final radiation (DGLAP evolution \cite{parisi_77}).  
In order to achieve a quantitative
description, within our HIJING/B\=B framework we
will show that combined effects of hard and soft sources of 
multiparticle production can reproduce the available data in the range
$0.02<\sqrt{s}< 20$ TeV only with a reduced dependence 
of the effective string tension on $\sqrt{s}$. 
We find that the data can be well reproduced
taking 
\begin{equation}
\kappa(s)= \kappa_{0} \,\,(s/s_{0})^{0.06}\,\,{\rm GeV/fm}\approx
Q_0\,Q_{\rm sat}(s),
\label{eq:kappa_sup}
\end{equation} 
where $\kappa_{0}$ = 1 GeV/fm is the vacuum string tension value, 
$s_{0}$ = 1 GeV$^2$ is a scale factor, and $Q_0$ is adjusted
to give $\kappa = 1.88$ GeV/fm at the RHIC energy $\sqrt{s}=0.2$ TeV.
Our phenomenological $\kappa(s)$ is compared to 
$Q_{\rm sat}^2(s)$  in Fig.~\ref{fig:kappa_vs_cgc}, where 
$\kappa$ = 1.40 GeV/fm at $\sqrt{s}$ = 0.017 TeV increases to
$\kappa$ = 3.14 GeV/fm at $\sqrt{s}$ = 14 TeV.     

\begin{figure} [h!]
\centering
\includegraphics[width=0.9\linewidth,height=6.0cm]{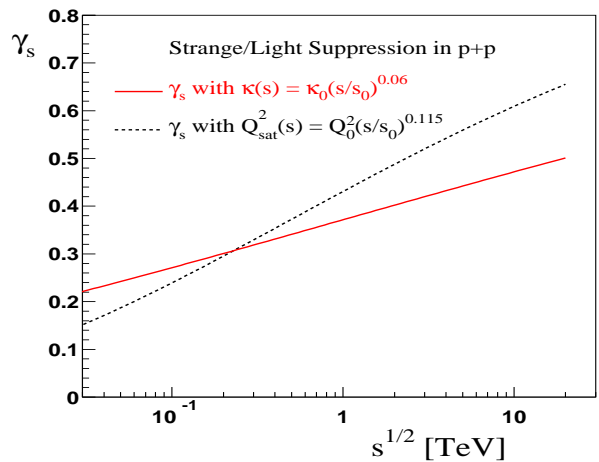}
\vskip 0.5cm\caption[supression of s quark s/u energy dependence] {\small 
(Color online) Energy dependence of the strange to light quark
 suppression factors, $\gamma_s = s/u$, using 
 $\kappa(s)$ from Eq.~\ref{eq:kappa_sup}  and using 
$Q_{\rm sat}^2(s)$ from Ref.~\cite{McLerran:2010ex} are compared.
(Note that this Schwinger suppression factor is not used
in the CGC model).
\label{fig:MG-fig1s_u}
}
\end{figure}

The energy dependence of the string tension leads to a variation 
of the diquark/quark suppression factors,
as well as the enhanced intrinsic transverse momentum $k_T$.  
These include  i) the ratio of production rates of  
diquark-quark to quark pairs (diquark-quark suppression factor),  
$\gamma_{{\rm qq}} = P({\rm qq}\overline{{\rm qq}})/P(q\bar{q})$;
ii) the ratio of production rates of strange 
to non-strange quark pairs (strangeness suppression factor), 
$\gamma_{s} = P(s\bar{s})/P(q\bar{q})$;
iii) the extra suppression associated with a diquark containing a
strange quark compared to
the normal suppression of strange quark ($\gamma_s$),
$\gamma_{us} = (P({\rm us}\overline{{\rm us}})/P({\rm
  ud}\overline{{\rm ud}}))/(\gamma_s)$;
iv) the suppression of spin 1 diquarks relative to spin 0 ones
(apart from the factor of 3 enhancement of the former based on
counting the number of spin states), $\gamma_{10}$; and 
v) the (anti)quark ($\sigma_{q}'' = \sqrt{\kappa/\kappa_0} \cdot \sigma_{q}$)
and  (anti)diquark ($\sigma_{\rm {qq}}'' = \sqrt{\kappa/\kappa_0}
\cdot {\it f} \cdot \sigma_{{\rm qq}}$) Gaussian  width.
As an example we plot in Fig.~\ref{fig:MG-fig1s_u} 
the energy dependence of the suppression factor $\gamma_{s}= s/u$, 
when the string tension values $\kappa(s)$ are taken from 
Eq.~\ref{eq:kappa_sup}
in comparison with the values 
predicted using $Q_{\rm sat}^2$ (Eq.~\ref{eq:larry10} 
from the CGC model fit \cite{McLerran:2010ex}). 

The contributions of multiple jets to the multiplicity distributions
in $pp$ and $p\bar{p}$ collisions have been studied in detail
in Ref.~\cite{wang_prd43_91}. Within the HIJING model, 
one assumes that nucleon-nucleon
collisions at high energy can be divided into soft and
hard processes with at least one pair of jet production with
$p_{T}>p_{0}$. A cut-off scale $p_0$ in the transverse momentum
of the final jet production has to be introduced below which the 
interaction is considered non-perturbative and can be characterized by
a finite soft parton cross section $\sigma_{\rm soft}$.
The inclusive jet cross section $\sigma_{\rm jet}$ at leading order
(LO) \cite{Eichten:1984eu} is 
\begin{equation}
\label{eq:sigma_jet}
 \sigma_{jet}=\int_{p_0^2}^{s/4}\mathrm{d}p_T^2\mathrm{d}y_1\mathrm{d}y_2 
 \frac{1}{2}\frac{\mathrm{d}\sigma_{jet}}{\mathrm{d}p_T^2
 \mathrm{d}y_1 \mathrm{d}y_2},
\end{equation}
where,
\begin{equation}
 \frac{\mathrm{d}\sigma_{jet}} {\mathrm{d}p_T^2 \mathrm{d}y_1 
\mathrm{d}y_2} =K \sum_{a,b}x_1f_a(x_1,p_T^2)x_2f_b(x_2,p_T^2) 
\frac{\mathrm{d}\sigma^{ab}(\hat{s},\hat{t},\hat{u})}{\mathrm{d}\hat{t}}
\end{equation}
depends on the parton-parton cross section $\sigma^{ab}$ and parton 
distribution functions (PDF) $f_a(x,p_T^2)$. The summation runs over all 
parton species; $y_1$ and $y_2$ are the rapidities of the scattered
partons; $x_1$ and $x_2$ are the light-cone momentum 
fractions carried by the initial partons.
The factor $K\approx2$ accounts for 
the next-to-leading order (NLO) corrections to the leading order
jet cross section. 
In the default {\small HIJING} model (v.1.383), the  
Duke-Owens parameterization \cite{Duke:1983gd} of PDFs in nucleons is used.
With the Duke-Owens parameterization of PDFs, an energy independent
cut-off scale $p_0=$2 GeV/$c$ and a constant soft
parton cross section $\sigma_{soft}=57$ mb are sufficient to
reproduce the experimental data on total and inelastic cross sections 
and the hadron central rapidity density in $p+p(\bar{p})$ collisions 
\cite{wang1_91,wang2_92}.   
Our results have been obtained using 
 the same set of parameters for hard scatterings as in the latest 
version of HIJING (v1.383).

\section{Charged Particles}   

\subsection{Charged Hadron Pseudorapidity}

\begin{figure} [h!]

\centering

\includegraphics[width=0.9\linewidth,height=4.0cm]{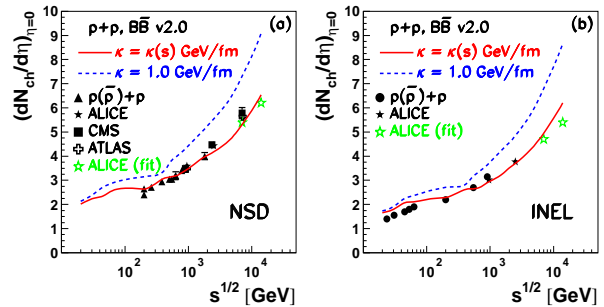}
\vskip 0.5cm\caption[NSD and Inelastic central dn/deta]
{\small (Color online) Comparison of {\small HIJING/B\=B v2.0} 
predictions for central charged particle pseudorapidity 
density in $pp$ and $p\bar{p}$
interactions for non-single-diffractive (NSD) (left panel) 
and inelastic (INEL)
(right panel) interactions as a function of c.m.s. energy.
The solid and dashed lines are the results
with and without SCF, respectively.
The data are from Refs. 
\cite{Khachatryan:2010xs,Khachatryan:2010us,alice:2009dt,Aamodt:2010ft,ua1_npb90,atlas_conf46,cdf1990_prd41,ua5_zfurp_c43,ua5_zfurp_c37,ua5_prep_154}
(left panel) and from
Refs.~\cite{alice:2009dt,npb_129_1977,ua5_zfurp_c37,ua5_prep_154}
(right panel). Only statistical error bars are shown. 
The open stars at 7 and 14 TeV (ALICE fit), 
are obtained by a power law fit to lower energy data from Refs. 
\cite{Aamodt:2010ft}, \cite{reygers_2010}. 
\label{fig:pp_dn_ch_0}
}

\end{figure}

Charged hadron multiplicity measurements are the first results 
of the LHC physics program. Data for $pp$ collisions were reported 
by the ALICE, ATLAS, and CMS collaborations
\cite{Khachatryan:2010xs,Khachatryan:2010us,alice:2009dt,Aamodt:2010my,Aamodt:2010pp,Aamodt:2010ft,Aad:2010rd,atlas_conf24,atlas_conf31,atlas_conf46,atlas_conf47}. 
The new data at mid-rapidity for non single diffractive interactions (NSD) 
and inelastic scattering (INEL)
are shown in Fig.~\ref{fig:pp_dn_ch_0}, which includes also 
similar results at lower energies. The
main result is an observed sizeable increase of the central pseudorapidity
density with c.m.s. energy.

The main contribution to the multiplicity comes from soft interactions 
with only a small component originating from
hard scattering of the partonic constituents of the proton.
In contrast to the higher $p_T$ regime, well described by pQCD,
particle production in soft collisions is generally modeled
phenomenologically to describe the different $pp$ scattering
processes: elastic scattering (el), single diffractive (SD), double
diffractive dissociation (DD), and inelastic non-diffractive
scattering (ND).
Experimentally, minimum bias events are a close approximation
of NSD interactions, {\it i.e.}, $\sigma_{\rm NSD}=\sigma_{\rm tot}-\sigma_{\rm
el}-\sigma_{\rm SD}$, where $\sigma_{\rm tot}$ is the total cross section.
The selection of NSD events is energy
dependent and differs somewhat for different experimental triggers.
The event selection of inelastic processes (INEL) 
includes SD interactions:
$\sigma_{\rm INEL}=\sigma_{\rm tot}-\sigma_{\rm el}$. 
The data therefore must be
corrected for the SD component, involving model dependent calculations.

The results of the model calculations 
for both NSD (left panel) and INEL (right panel) are also shown in
Fig.~\ref{fig:pp_dn_ch_0}. Solid lines depict results including 
the strong color field (SCF) effects
whereas the results without SCF effects are shown as dashed lines.
Without SCF effects the model strongly overestimates the central
charged particle density. 
The absolute value and energy increase of the
central rapidity density are well reproduced assuming the energy dependent
string tension given in Eq.~\ref{eq:kappa_sup}.
At higher LHC energies (2.36 and 7 TeV), a discrepancy
of 10-15 \% is observed.

As the colliding energy increases, the rate of
multiple parton interactions (MPI)
also increases, producing a rise in the central multiplicity.
The increase with energy in our phenomenology
is due to the interplay of the increased mini-jet production
in high colliding energy with SCF effects.
For an increase of strangeness suppression factors due to an
increase of string tension with energy ($\kappa=\kappa(s)$),
the model predicts a decrease of produced pions due to energy conservation.
Lower values of $\kappa(s)$ imply smaller values for strangeness
suppression factors, therefore a higher multiplicity of mesons
(mostly pions). 

\begin{figure} [t!]

\centering
\includegraphics[width=8.6cm,height=8.6cm]{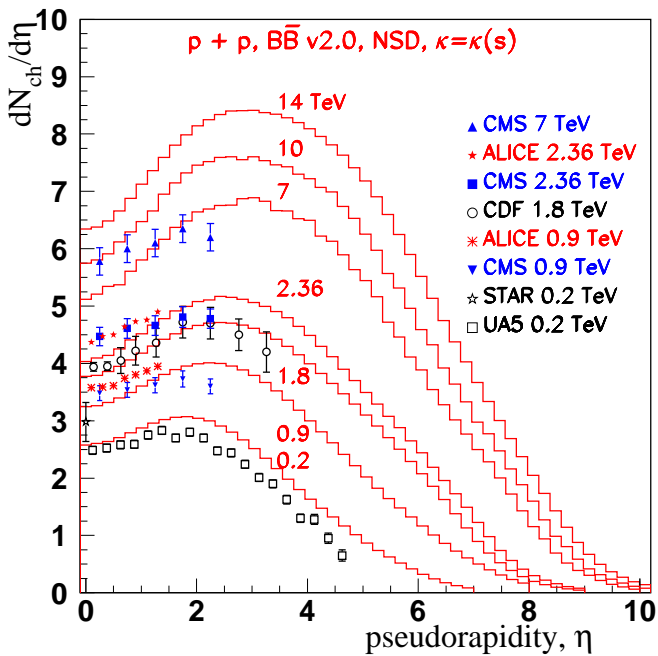}
\vskip 0.5cm\caption[pseudorapidity energy dependence] {\small
(Color online) Comparison of {\small HIJING/B\=B v2.0} predictions for
the pseudorapidity distribution of charged particles in 
$p$ + $p$ ($\bar{p}$) collisions at various c.m.s. energies. 
The solid histograms are the results with SCF and J\=J loops. 
The data are from Refs. \cite{ua5_zfurp_c33} (UA5), \cite{:2008ez} (STAR), 
\cite{cdf1990_prd41} (CDF), 
\cite{alice:2009dt,Aamodt:2010pp} (ALICE),
\cite{Khachatryan:2010xs,Khachatryan:2010us} (CMS).
Only statistical error bars are shown.
\label{fig:pp_dn_ch_eta}
}
\end{figure}

Changing the effective value $\kappa(s)$ = 2.89 GeV/fm to
$\kappa(s)$ = 2.0 GeV/fm results in an
increased multiplicity of 11\% at 7 TeV where the effect is greatest.
In addition, the multiplicity depends also on the value of the 
cut-off parameter $p_0$. Low values of $p_0$ imply high rates of 
parton-parton scattering
and hence high levels of particle multiplicity. Evidently for increasing
values of $p_0$ the opposite is expected.
At 7 TeV, where the effect is also the largest, changing $p_0$ by
$\pm \, 0.5$ GeV results
in a change of only $\mp \, 1.8\%$ for the central pseudorapidity density.

Data on the charged particle pseudorapidity distribution are also 
available over a limited $\eta$ range  
\cite{Khachatryan:2010xs,Khachatryan:2010us}
\cite{alice:2009dt,Aamodt:2010pp,Aamodt:2010ft}. 
These are presented in
Fig.~\ref{fig:pp_dn_ch_eta} where we also include results
obtained at 0.2 TeV by the UA5 \cite{ua5_zfurp_c33}  
and STAR \cite{:2008ez} collaborations, and with CDF results
\cite{cdf1990_prd41}
obtained at the Tevatron for $p\bar{p}$ collisions at 1.8 TeV.
Consistent with the discussion above, a scenario with SCF effects (solid
histograms) reproduces the measured multiplicity distributions well.
At all energies considered,
 theoretical calculations predict a central dip at mid-rapidity
that is consistent with the observations. At 0.9 and 2.36 TeV, the shape of the
distribution measured by the ALICE collaboration is very well reproduced, while
the CMS results show a much flatter distributions than calculated 
(also the case at 7 TeV). Data over a larger rapidity range are needed 
to determine the shape of the falling
density in the fragmentation region. For completeness,
predictions at 10 and 14 TeV, the higher LHC energies, are also shown.

\subsection{Transverse momentum spectra}

\begin{figure} [h!]

\centering

\includegraphics[width=8.6cm,height=8.6cm]{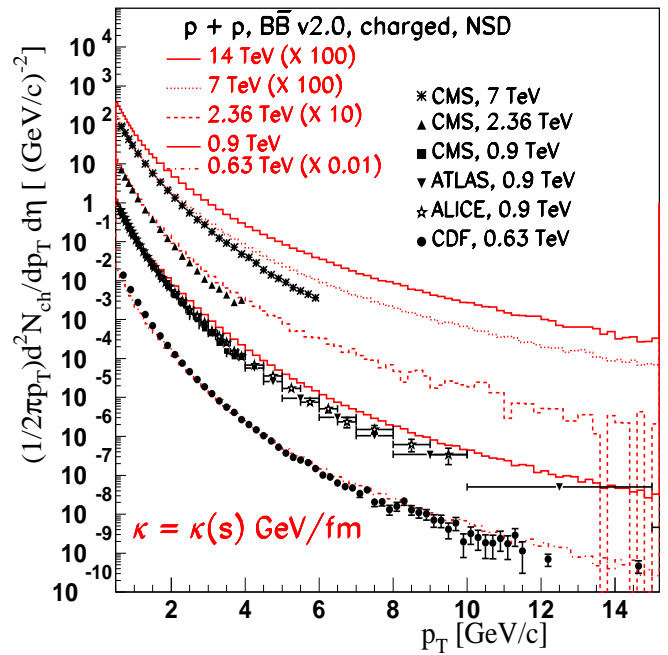}
\vskip 0.5cm\caption[transverse momentum distribution 0.63-14 TeV]
{\small (Color online) Comparison with data 
of {\small HIJING/B\=B v2.0} predictions of
charged-hadron transverse momentum distributions at LHC energies.
The calculated spectra include the combined effects of SCF and J\=J loops.
The histograms and data have been scaled for clarity by the factors indicated.
The data are from Refs.~\cite{cdf_prd65_2002} (CDF),
\cite{Aamodt:2010my} (ALICE), \cite{Aad:2010rd} (ATLAS),
\cite{Khachatryan:2010xs,Khachatryan:2010us} (CMS). 
Error bars include only the statistical uncertainties. 
\label{fig:pp_ch_pt_all}
}

\end{figure}

The measured transverse momentum distributions for NSD events 
over an energy range
$\sqrt{s}=0.63-7$ TeV are shown
in Fig.~\ref{fig:pp_ch_pt_all}. These recent measurements are
performed in the central rapidity region and
cover a wide $p_T$ range ($0.15 < p_T < 10$ GeV/c), where
both hard and soft processes are expected to contribute.
The data of ATLAS and CMS are measured in larger pseudorapidity
intervals ($|\eta| < 2.5 $). In contrast, ALICE and CDF measurements are in a
very central region ($|\eta| < 0.8$ and $|\eta| < 1.0$, respectively). The
calculation takes into account the difference in acceptance, but as can be
concluded from Fig.~\ref{fig:pp_dn_ch_eta}, this difference in
pseudorapidity range has a negligible
effect on the measured cross section.

The model calculations including SCF effects
describe the data well at $\sqrt{s}=0.63$
TeV, but lead to a harder spectrum that observed at higher energy. The model
gives a fair description of the spectral shape at low $p_T$
but overestimate the data at high $p_T$.
The discrepancy is highest (up to a factor of three) at $\sqrt{s}= 7$
TeV. We do not understand the source of this discrepancy and await 
higher $p_T$ data to draw a firm conclusion.
However, in our phenomenology this could indicate 
that jet quenching, {\it i.e.}
suppression of high $p_T$ particles like that observed at RHIC energies
in nucleus-nucleus collisions, could also appear in $pp$ collisions
in events with large multiplicity.
This overestimation of high $p_T$ yield leads also to a similar overestimation
of the mean transverse
momentum ($<p_T>$) and of the correlation of mean  $<p_T>$ as a function of
N$_{\rm ch}$, which we do not discuss here.

Within our model we generate events with different numbers of mini-jets up to
some maximum values.
High numbers of mini-jets lead to higher multiplicity events.
For events with ten mini-jets (the maximum assumed in our calculation),  
the central charged particle
pseudorapidity density could increase up to $\approx$ 20
and the total multiplicity could be greater than 150 at Tevatron
energy (1.8 TeV).
The measurements of two particle correlations over the entire azimuth 
could reveal the jet structure
related to high $p_T$ particles \cite{Tannenbaum:2010ab}.
The study of these correlations in events
with high multiplicity
could help us to draw a firm conclusion with regard to a
possible jet quenching phenomenon in $pp$ collisions.

\section{Identified Particle Spectra and Ratios}

\subsection{Baryon-to-Meson ratio}

The $pp$ single particle inclusive $p_T$ spectra measurements
are important for understanding collision dynamics, since
the various particles show different systematic behavior,
as observed at RHIC energy \cite{Tannenbaum:2010ab}. 
Detailed theoretical predictions for single inclusive hadron
production (including strange hyperons) are discussed in this section.
Baryon-to-meson ratios (B/M)
are experimental observables that can be used at the LHC   
for investigating multi-parton interactions 
and helping to understand 
the underlying physics \cite{Hippolyte:2006ra,Hippolyte:2009xz,Ricaud:2010ay}.

\begin{figure} [h!]
\centering
\includegraphics[width=0.9\linewidth,height=4.0cm]{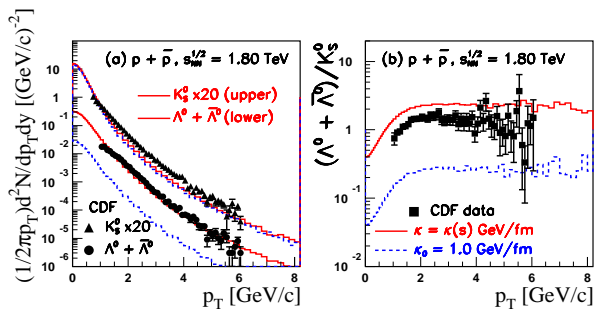}
\vskip 0.5cm\caption[strange baryon/mesons at 1.8 TeV -CDF data] 
{\small (Color online) {\small HIJING/B\=B v2.0}
  predictions at $\sqrt{s}$ = 1.8 TeV of baryon and mesons transverse momentum 
   at mid-rapidity ($-1 < y < 1$) (a) and their ratios 
for single strange particles ($ \Lambda^0 + \bar{\Lambda}^0 $)/$K_S^0$ (b) 
in minimum bias $p\bar{p}$ collisions.
The solid and dashed lines have the same meaning as in
Fig.~\ref{fig:pp_dn_ch_0}.  
Experimental results at $\sqrt{s}$ = 1.8 TeV
are from Ref.~\cite{Acosta:2005pk} (CDF Collaboration).  
Error bars include only the statistical uncertainties. The 
ratios (b) have been calculated by us, dividing the spectra 
in the panel (a).  
\label{fig:pp_ks0_l0_18}
}
\end{figure}

Unexpectedly high B/M ratios observed in $A+A$ collisions 
have been discussed in terms of recombination and coalescence mechanisms
\cite{Fries:2003kq,Greco:2003mm,Hwa:2002tu}.
Such high ratios at intermediate $p_T$ were also
reported in $pp$ collisions at RHIC \cite{Abelev:2006cs} and at the Tevatron
\cite{Acosta:2005pk}. In $pp$ collisions, however,
a coalescence/hadronization scenario  is not
favored due to low phase space density in the final state.
Our {\small HIJING/B\=B} model, with SCF effects included,
provides an alternative dynamical
explanation of the heavy-ion data at RHIC energies.
We have shown that the model also
predicts an increasing yield of (multi)strange particles, thereby 
better describing the experimental data \cite{prc75_top07}. 

\begin{figure} [h!]
\centering
\includegraphics[width=0.9\linewidth,height=4.0cm]{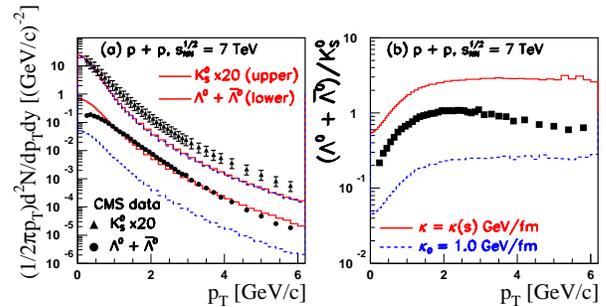}
\vskip 0.5cm\caption[strange baryon/meson ratio at 7 TEV -CMS data] 
{\small (Color online) 
Predictions of the {\small HIJING/B\=B v2.0}
 model for baryon and mesons transverse momentum 
 in the rapidity range $-2 < y < 2$ (a) and their ratio (b) 
for minimum bias $p\bar{p}$ collisions at $\sqrt{s}$ = 7.0 TeV.
The solid and dashed lines have the same meaning as in
Fig.~\ref{fig:pp_dn_ch_0}. The experimental results (a) 
are from Ref.~\cite{cms_strang_2010}(CMS Collaboration) .  
Error bars include only the statistical uncertainties. The 
ratios (b) have been calculated by us, dividing the spectra 
in the panel (a). 
\label{fig:pp_ks0_l0_7}
}
\end{figure}

Figure~\ref{fig:pp_ks0_l0_18} (b) shows a comparison of model 
predictions with CDF experimental data \cite{Acosta:2005pk}
of the strange baryon-to-meson ratio
$(\Lambda^0 + \bar{\Lambda}^0)/K_S^0$ at 1.8 TeV.
The particle $p_T$ spectra are shown in the panel (a).
The measured ratio is fairly well described within our
phenomenology. The larger string tension parameterization results 
in a predicted increase of the ratio $(\Lambda^0 + \bar{\Lambda}^0)/K_S^0$ 
by a factor of $\approx$ 10 at the Tevatron energy.
The $p_T$ spectra show that the increased ratio is due almost entirely to an
increase of the $\Lambda$ cross section that is well described 
with $\kappa(s)= \kappa_{0} \,\,(s/s_{0})^{0.06}\,\,{\rm GeV/fm}$.
The model underestimates the kaon production by 15-25 \%.

The model predictions at 7 TeV, currently the maximum energy 
where there are data
\cite{cms_strang_2010}, are shown in Fig.~\ref{fig:pp_ks0_l0_7} (b). 
The model gives a good description of
hyperon production ($\Lambda^0 + \bar{\Lambda}^0 $), for which
an increase by a factor of ten is still predicted if SCF effects
are considered. However, our calculations underestimate by approximately
a factor of two the yield of $K_S^0$ and, as a consequence, result in 
a higher ratio than that observed. This result needs further
investigation (theoretical and experimental) on heavy flavour 
production at this energy. 
We note that the models PYTHIA
\cite{Sjostrand:2007gs,Sjostrand:2006za} 
and Energy-conserving Partons Off-shell remnants and Spliting of
partons ladders (EPOS) \cite{Werner:2009zz,Werner:2010zz} 
cannot reproduce the observed high
B/M ratio (see Fig. 6 and Fig. 7 from Ref.~\cite{Hippolyte:2009xz}).
The preliminary data reported by the CMS collaboration
indicate high hyperon yield. Comparisons with PYTHIA results show that
this model significantly
underestimates the hyperon yields in $pp$ collisions
at 0.9 and 7 TeV \cite{cms_strang_2010}.

The strange particle ratios could also be the manifestation of new collective
phenomena.
In the EPOS model such an increase is obtained if the production 
of a {\em mini-plasma} is
considered in $pp$ collisions \cite{armesto_jpg08}, \cite{Werner:2010zz}.
If confirmed by future measurements,
the study of these observables could open a perspective on new physics in
$pp$ interactions.

\begin{figure} [h!]
\centering
\includegraphics[width=0.9\linewidth,height=4.0cm]{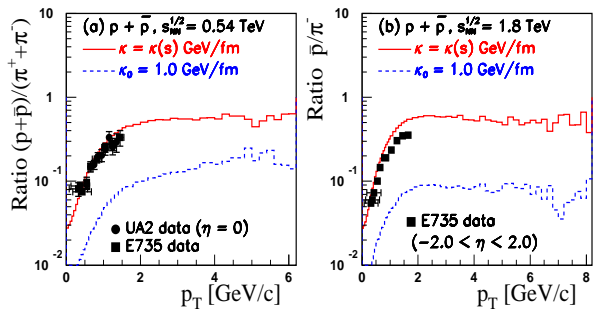}
\vskip 0.5cm\caption[pbar/pi- ratio at 0.54 Tev and 1.8 TeV] 
{\small (Color online) 
Comparison of {\small HIJING/B\=B v2.0} predictions 
with data on the
non-strange baryon over meson ratios from minimum bias events in the
rapidity range $|y| < 2$.
The solid and dashed lines have the same meaning as in
Fig.~\ref{fig:pp_dn_ch_0}.  
Experimental results for $|y| < 2$ at $\sqrt{s}$ = 0.54 TeV (left panel)
and at $\sqrt{s}$ = 1.8 TeV (right panel)
are from Ref.~\cite{prd_48_93_e735} (E735 Collaboration).
The results at mid-rapidity at $\sqrt{s}$ = 0.54 TeV (left panel) are
from Ref.~\cite{pl_122b_93_ua2} (UA2 Collaboration).   
Error bars include only the statistical uncertainties.  
\label{fig:pp_pi_pbar_18}
}
\end{figure}

Similar conclusions are obtained from the study of the
proton/pion ($p/\pi^{+}$, $\bar{p}/\pi^{-}$) ratios where data exist at lower
energies \cite{prd_48_93_e735,pl_122b_93_ua2}. These data, shown in
Figure~\ref{fig:pp_pi_pbar_18}, are limited to low $p_T < 2$ GeV/c.
Adding SCF effects results in a very sizable increase of the ratio and
our calculations  provide a good
description of the data in the measured range.  
However, as the calculations indicate, to draw a final conclusion 
measurements at intermediate and high $p_T$ are needed.

\begin{figure} [h!]
\centering
\includegraphics[width=0.9\linewidth,height=6.0cm]{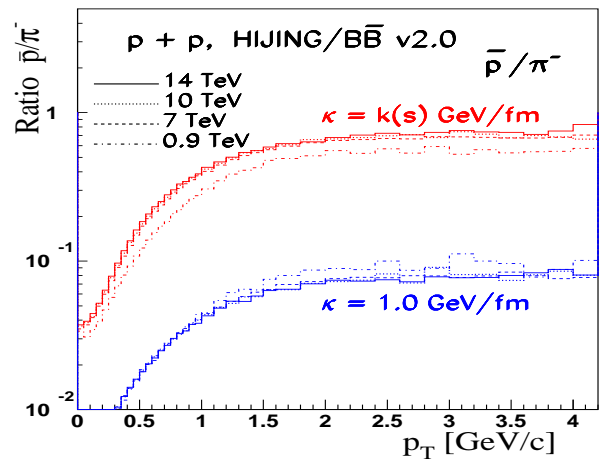}
\vskip 0.5cm\caption[ratio pbar/pi- from 0.9 TeV - 14 TeV] 
{\small (Color online) 
Predictions of the {\small HIJING/B\=B v2.0} model
of non-strange baryon over meson ratios ($\bar{p}/\pi^-$) 
for minimum bias events at mid-rapidity at LHC energies.
The upper curves correspond to calculations
which include  the effects of SCF and J\=J loops.
The lower curves corresponds to calculations without SCF effects.
The results are at 0.9 TeV (dot dashed-histograms),
at 7 TeV (dashed histograms), at 10 TeV (dotted histograms) 
and at 14 TeV ( solid histograms). 
 \label{fig:pp_pi_pbar_14}
}
\end{figure}

To the extent that the LHC experiments are able 
to identify hadron species, such data
will provide vital input to validate this interpretation.
The model predictions at LHC
energies for the $p_T$ dependence of the $\bar{p}/\pi^-$ ratio
are shown in Fig.~\ref{fig:pp_pi_pbar_14}.
An enhancement up to the highest LHC
energy and a weak energy dependence, with a saturation that sets in
for a c.m.s. energy $\sqrt{s} > 2.36$ TeV, is predicted. 
Note that preliminary data at 0.9 TeV reported by the ALICE collaboration 
for $p_T$ spectra of pions ($\pi^+$) and protons ($p$) 
\cite{Antonioli:2010eq} cover to $p_T < 2.5$ GeV/c. 
The model results, with SCF effects included (dot-dashed histogram 
in Fig.~\ref{fig:pp_pi_pbar_14})
are consistent with a $p/\pi^+$ ratio derived from the spectra
reported by ALICE at 0.9 TeV in Ref.~\cite{Antonioli:2010eq}.

In our approach, the dynamical
mechanism that leads to such high values of B/M ratios  
is SCF appearing at the initial stage of the interaction.
The SCF mechanism strongly modifies the fragmentation processes
(strangeness suppression factors) and thus results in a huge increase of
(strange)baryons. This interpretation is also supported by more sophisticated
theoretical calculations, in a scenario in which a 
time-dependent pulse for the initial strength of
the color field is considered \cite{Skokov:2007gy,Skokov:2009zz}.
The large enhancement of the baryon-to-meson ratios 
demonstrates that SCF
could play an important role in multiparticle production in $pp$
collisions at LHC energies and that high energy density fluctuations
can reach very high densities, potentially comparable to those reached 
in central Au + Au collisions at RHIC energies \cite{Giovannini:2003rv}.

\subsection{Baryon-anti-baryon asymmetry}

From the study of the baryon-anti-baryon asymmetry one can learn about the
mechanism of baryon number transport.
Baryon production via the conventional default quark-diquark mechanisms in
the Lund string model are known to be inadequate even in $e^++e^-$ 
phenomenology.
This is one of the main reasons for
our continued exploration of alternative baryon junction mechanisms.
The details of the new implementation of J\=J loops in
{\small HIJING/B\=B v2.0} are described in
Ref.~\cite{prc75_top07} Sec II A.
In {\small HIJING/B\=B v2.0} the main two mechanisms for baryon
production are quark-diquark ($q-{\rm qq}$) string fragmentation and junction
anti-junction (J\=J) loops \cite{Ripka:2003vv}
in which baryons are produced approximately in pairs. In a junction loop a
color flux line splits at some intermediate point
into two flux lines at one junction and then the flux lines fuse back
at an anti-junction somewhere further
along the original flux line.
The distance in rapidity between these points is chosen
via a Regge distribution ~\cite{prc75_top07}.
We assume that, out of the non single diffractive nucleon-nucleon
(NN) interaction cross section ($\sigma_{\rm NSD}$),
a fraction
$f_{J\bar{J}}=\sigma_{J \bar{J}}/(\sigma_{\rm INEL}-\sigma_{\rm SD})$
of the events excite a junction loop.
The probability that the incident baryon
has a J\=J loop in p(A)+A collisions after $n_{\rm hits}$
simulated binary collisions is given by

\begin{equation}
P_{J\bar{J}}=1-(1-f_{J\bar{J}})^{n_{hits}},
\end{equation}

where $\sigma_{J \bar{J}}$ = 17 mb,
$\sigma_{\rm SD}$ is parametrized in the model,
and the total inelastic nucleon-nucleon cross sections
are calculated \cite{mik_wang_94}.
These cross sections imply that a junction loop occurs in $pp$ collisions
with a rather high probability (at RHIC energy $f_{J\bar{J}}\approx 0.5$
for $\sigma_{\rm INEL}$ = 42 mb). Taking a
constant value for  $\sigma_{J \bar{J}}$ results in a decrease
with energy of the probability $P_{J\bar{J}}$, due to a faster
increase with energy of $\sigma_{\rm INEL}$ relative to
$\sigma_{\rm SD}$.
The actual probability is modified also
by string fragmentation processes for which we consider a threshold cutoff mass
$M_c= 6$ GeV/c$^2$ in order to have enough kinematic phase space
to produce B\=B pairs.
We have investigated the sensitivity of the results to the value
of parameters  $\sigma_{J \bar{J}}$ and $M_c$ and found no significant
variation on pseudorapidity distributions of charged particles
for 15 mb $< \sigma_{J \bar{J}} <$ 25 mb and for
4 GeV/c$^2$ $< M_c <$ 6 GeV/c$^2$. 

Baryon number transport is quantified in terms of the rapidity loss
($\delta y_{\rm loss} = \delta y_{\rm beam}-\delta y_{\rm baryon}$,
where $y_{\rm beam}$ and $y_{\rm baryon}$ are the rapidity of incoming
beam and outgoing baryon, respectively) and has been discussed
within our model phenomenology for $A+A$ collisions
in Refs.~\cite{prc72_top05,prc75_top07,armesto2_08}.
It was shown that {\small HIJING/B\=B v1.0}
overestimate the stopping power and 
give a mild energy dependence of net-baryons at mid-rapidity.
The energy dependence of net-baryons at mid-rapidity
per participant pair within  {\small HIJING/B\=B v2.0}
is proportional to $(s/s_0)^{-1/4 +\Delta/2}$,
similar to the dependence predicted in Ref. \cite{Kharzeev:1996sq},
with the assumption that J\=J is the dominant mechanism.
This dependence is obtained assuming the following 
parameters \cite{armesto2_08}:
$s_0 = 1$ GeV$^2$ (the usual parameter of Regge theory),
$\alpha(0)=1/2$ (the reggeon intercept of the trajectory), and
$\alpha_{P}(0) = 1 + \Delta$ (where $\Delta \approx 0.01$) for the
pomeron intercept.

\begin{figure} [h!]
\centering

\includegraphics[width=0.9\linewidth,height=4.0cm]{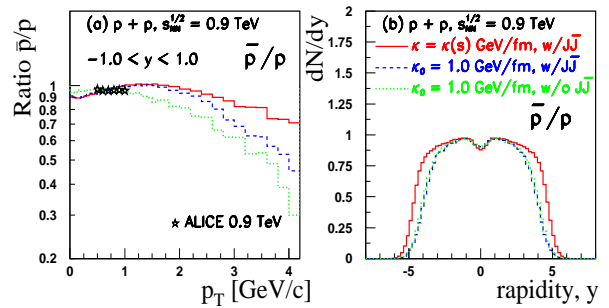}
\vskip 0.5cm\caption[pbar/p rapidity and pt dependence at 0.9 TeV] 
{\small (Color online) {\small HIJING/B\=B v2.0}
 predictions $p_T$ distributions (a) at mid-rapidity for 
the $\bar{p}/p$ ratio  at $\sqrt{s}$ = 0.9 TeV; 
rapidity distributions are shown in part (b). 
The results are from three possible scenarios:
without contributions from SCF and J\=J loops (dotted histograms),
including only the effect of  J\=J loops (dashed histograms)
and including both effects (SCF and J\=J loops) (solid histograms).
The data are from Ref.~\cite{Aamodt:2010dx} (ALICE Collaboration).
Only statistical errors are shown.
 \label{fig:pp_rap_pbar_pt}
}
\end{figure}

Recently, the ALICE Collaboration reported results \cite{Aamodt:2010dx}
on mid-rapidity anti-proton-to-proton ratio in $pp$ collisions at
$\sqrt{s}=0.9$ and 7 TeV and equivalently the proton-anti-proton
asymmetry, $A = (N_{p} - N_{\bar{p}})/(N_{p} + N_{\bar{p}})$. These data
could be used to constrain Regge inspired model
descriptions of baryon asymmetry.
The authors state that, within statistical errors, the 
observed $\bar{p}/p$ ratio shows no dependence
on transverse momentum or rapidity in the limited measured acceptance
($-0.8<y<0.8$; $0<p_T< 1 $ GeV/c).
In Fig.~\ref{fig:pp_rap_pbar_pt} (a) are compared the
{\small HIJING/B\=B v2.0} model predictions with the
published data at 0.9 TeV.
Our model predicts negligible dependence on $p_T$ for $0<p_T<2$ GeV/c,
and a slight $p_T$ dependence at higher $p_T$, where both
effects (J\=J loops and SCF) could contribute. The full
calculation is shown by the solid histogram. The dotted line is the prediction
without J\=J loops and SCF effects while the dashed line includes the effect of
J\=J loops only.
Over the measured ranges the rapidity distribution of the ratio $\bar{p}/p$ 
distribution is not sensitive to the various scenarios presented.
The scenario with combined effects results in a wider rapidity 
distribution at $y > 3 $ (b). The narrow structure observed near 
$y = 0$ has no physical significance and we believe that it is likely due to 
a numerical artifact of our current implementation of fragmentation
scheme.

\begin{figure} [h!]
\centering

\includegraphics[width=0.9\linewidth,height=4.0cm]{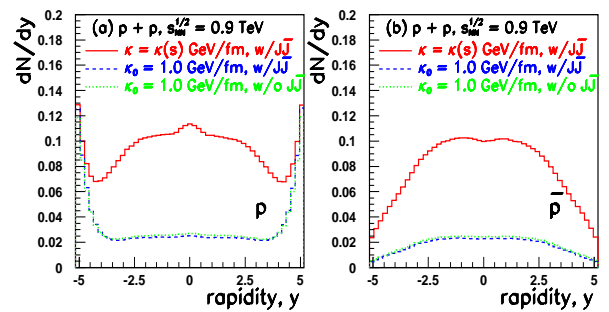}
\vskip 0.5cm\caption[p and pbar at 0.9 TeV rapidity distributions] 
{\small (Color online) 
Predictions of the {\small HIJING/B\=B v2.0} model for rapidity 
distributions of protons (a) and anti-protons (b)
at $\sqrt{s}$ = 0.9 TeV. The histograms have the same meaning as 
in Fig.~\ref{fig:pp_rap_pbar_pt}.
\label{fig:pp_p_pbar_rap}
}
\end{figure}

Separate proton and anti-proton rapidity distributions are, 
however, much more
sensitive to SCF effects, as seen in Fig.~\ref{fig:pp_p_pbar_rap}.
The model predicts a substantial increase (by a factor of $\approx 5$)
for p($\bar{p}$), when SCF are taken into account.
Due to the high cutoff mass $M_c=6$ GeV/c$^2$ 
the effect of J\=J loops is very small over the entire rapidity region.
Analysis of the 7 TeV data leads to similar conclusions, but 
shows less effect at high $p_T$.

Over the measured range 
the feed-down-corrected data for the $\bar{p}/p$ ratio rises from
$0.957 \pm 0.006$ at 0.9 TeV to $0.991 \pm 0.005$ 
at 7 TeV \cite{Aamodt:2010dx}.
Although the measured mid-rapidity ratio is close to unity
there is a small but significant excess of protons over
anti-protons corresponding to an asymmetry of
$A=0.022 \pm 0.003$ and $A=0.005 \pm 0.003$ at $\sqrt{s} = 0.9$ and 7 TeV,
respectively.
Within our model using a Regge intercept of $\alpha_0$ = 1/2 
results in a proton-anti-proton asymmetry 
$A_{\rm model}$ = 0.032 (0.002) at 0.9 TeV (7 TeV). 
The result overestimates the value of asymmetry at 0.9 TeV. 
These values show that
a fraction of the baryon number associated with the beam particles
is transported over rapidity intervals of less than seven units.
However, a better understanding of the baryon transport would require measuring
the rapidity dependence of the asymmetry over a larger range.

In Ref.~\cite{prc75_top07} we discuss the predicted rapidity
correlation length for production of baryon-anti-baryon pairs within our model.
The predicted rapidity correlation length ($1-\alpha(0))^{-1}$ depends upon the
value of the Regge intercept $\alpha(0)$ \cite{miklos_99}.
A value of $\alpha(0) \simeq 0.5$ 
leads to rapidity correlations in the range $|y_B-y_{\bar{B}}| \sim 2$,
while a value $\alpha(0) \simeq 1.0$ \cite{kopelovic99}
is associated with infinite range rapidity correlations.
This kind of analysis and future measurements at
LHC energies will help us
to determine better the specific parameters characterizing J\=J loops
used in the calculations of baryon number transport and/or
baryon-anti-baryon asymmetry  
while helping to establish the validity of Regge inspired
models.

\section{Summary and Conclusions}

We have studied the influence of strong longitudinal color fields
and of possible multi-gluon dynamics ({\em gluon junctions})
in particle production in $pp$ collisions, 
with a focus on RHIC, to Tevatron, and LHC energies.
We have investigated a set of observables
sensitive to the dynamics of the collisions,
covering both longitudinal and transverse degree of freedom.
A detailed comparison with newly
available experimental data from the LHC has been performed.

We found that the inclusion of the multiple minijet source 
limits the growth of the string tension $\kappa(s)$ to be approximately 
only linear as a function of saturation scale $Q_{\rm sat}$ 
(Eq.~\ref{eq:kappa_sup}), in contrast to recent
approaches \cite{McLerran:2010ex} where $\kappa(s)$ 
scales as $Q_{\rm sat}^2$ (Eq.~\ref{eq:larry10}). 
The reason for this is that in the CGC model the collinear factorized minijet 
mechanism is suppressed by geometric scaling to much higher $p_T$. 
Future measurements at LHC energies ($\sqrt{s}$ = 7 and 14 TeV),
extended to high $p_T$, will help to clarify the validity of  
this mechanism.

We have shown that SCF could play an important role in
particle production at mid-rapidity in $pp$ collisions.
Our calculations show that high-energy density fluctuations in $pp$
collisions at LHC can reach densities comparable
to those reached in central nuclear ($A+A$) collisions at RHIC.
A large enhancement of the (strange)baryon-to-meson ratios
that persists up to the highest LHC energy
can be explained as an effect of SCF
that appears at the initial stage of the interaction.
The mechanisms of hadron production
are very sensitive to the early phase of the
collisions, when fluctuations of the color field strength are highest.
Strong Color Field effects are modeled
by varying the effective  string tension that controls the
$q\bar{q}$ and ${\rm qq}\overline{\rm qq}$ pair creation rates
and strangeness suppression factors.
SCF, therefore, may modify the fragmentation processes 
with a resultant huge increase of (strange)baryons.

We show that both J\=J loops and SCF effects could play an important role in
baryon (anti-baryon) production at mid-rapidity in $pp$ collisions at
LHC energies. Introducing a new J\=J loop algorithm
in the framework of  {\small HIJING/B\=B} v2.0
leads to a consistent and significant improvement in the description
of the recent experimental results for proton-anti-proton
and for baryon-anti-baryon asymmetry in comparison to the older versions
HIJING/B or HIJING/B\=B v1.0 \cite{miklos_99}.
We have shown that baryon number transport is suppressed for
$\delta y > 7$, a result that is confirmed by recent ALICE measurements
 \cite{Aamodt:2010dx}.

The present study is limited to
the effect of initial state baryon production via possible junction dynamics
in strong fields. It would be very useful 
to consider a generalization of back reaction effects \cite{Tanji:2008ku}
to the case not only of pair production relevant for mesons
but to the more difficult
three string junction configurations needed to describe baryon production.

A greater sensitivity to SCF effects is expected
also for open charm and bottom production \cite{top_prl2009}.
In particular, measurements of rapidity and $p_T$ distributions
for particles involving charm and bottom quark,
would provide an important test of the relevance of SCF fluctuations,
helping us to determine values of the suppression
factors $\gamma_{Q\bar{Q}}$ (where Q = qq, $s$, $c$, $b$), which  
have strong dependence on the main parameters of QCD
(the constituent and current quark masses) and on the system size.
Even though the success of this procedure has been clearly illustrated here,
a fuller understanding
of particle production and especially of (multi)strange particles in
ultra-relativistic $pp$ collisions at the LHC remains an exciting open question,
and will continue to challenge many theoretical ideas.

\section{Acknowledgments}
\vskip 0.2cm 

{\bf Acknowledgments:} We thank S. Das Gupta and S. Jeon for useful 
discussions and continued support. We thank P. Levai for helpful
discussions and suggestions throughout this project. 
VTP acknowledges computer facilities at Columbia
University, New York, where part of these calculations were performed.
This work was supported by the Natural Sciences and Engineering 
Research Council of Canada.  
This work was also supported by the Division of Nuclear Science, 
of the U. S. Department of Energy under Contract No. DE-AC03-76SF00098 and
DE-FG02-93ER-40764.

\end{document}